\documentclass[aps,11pt,prd,groupedaddress,nofootinbib,notitlepage,eqsecnum,preprintnumbers]{revtex4-2}
\usepackage[utf8]{inputenc}
\usepackage{hyperref}
\usepackage[dvipsnames]{xcolor}
\usepackage{graphicx}
\usepackage{amsmath,amssymb}
\usepackage{bm}
\usepackage{comment}
\usepackage[shortlabels]{enumitem}
\usepackage{overpic}
\usepackage{ulem}

\usepackage[framemethod=default]{mdframed}
\newmdenv[skipabove=7pt,
skipbelow=7pt,
rightline=false,
leftline=false,
topline=false,
bottomline=false,
backgroundcolor=gray!10,
linecolor=gray,
innerleftmargin=5pt,
innerrightmargin=5pt,
innertopmargin=5pt,
innerbottommargin=5pt,
leftmargin=0cm,
rightmargin=0cm,
linewidth=4pt]{eBox}

\begin{document}

\title{Parametric resonance of gravitational waves in general scalar-tensor theories}

\author{\textsc{Yi-Fu Cai}$^{a,b}$}
    \email{yifucai@ustc.edu.cn}
\author{\textsc{Guillem Domènech}$^{c}$}
    \email{guillem.domenech@itp.uni-hannover.de}
\author{\textsc{Alexander Ganz}$^{c,d}$}
    \email{{alexander.ganz}@{itp.uni-hannover.de}}
\author{\textsc{Jie Jiang}$^{b,e,f}$}
    \email{jiejiang@pusan.ac.kr}
\author{\textsc{Chunshan Lin}$^d$}
    \email{chunshan.lin@uj.edu.pl}
\author{\textsc{Bo Wang}$^{a,b}$}
    \email{ymwangbo@ustc.edu.cn}
\affiliation{$^a$ Deep Space Exploration Laboratory/School of Physical Sciences, University of Science and Technology of China, Hefei, Anhui 230026, China}
\affiliation{$^b$ Department of Astronomy, School of Astronomy and Space Science, University of Science and Technology of China, Hefei, Anhui 230026, China}
\affiliation{$^c$ Institute for Theoretical Physics, Leibniz University Hannover, Appelstraße 2, 30167 Hannover, Germany.}
\affiliation{$^d$ Faculty of Physics, Astronomy and Applied Computer Science, Jagiellonian University, 30-348 Krakow, Poland}
\affiliation{$^e$ Center for Cosmological Constant Problem, Pusan National University, Busan 46241, Republic of Korea}
\affiliation{$^f$ Department of Physics, Pusan National University, Busan 46241, Republic of Korea}

\begin{abstract}
    Gravitational waves offer a potent mean to test the underlying theory of gravity. In general theories of gravity, such as scalar-tensor theories, one expects modifications in the friction term and the sound speed in the gravitational wave equation. In that case, rapid oscillations in such coefficients, e.g. due to an oscillating scalar field, may lead to narrow parametric resonances in the gravitational wave strain. We perform a general analysis of such possibility within DHOST theories. We use disformal transformations to find the theory space with larger resonances, within an effective field theory approach. We then apply our formalism to a non-minimally coupled ultra-light dark matter scalar field, assuming the presence of a primordial gravitational wave background, e.g., from inflation. We find that the resonant peaks in the spectral density may be detectable by forthcoming detectors such as LISA, Taiji, Einstein Telescope and Cosmic Explorer. 
\end{abstract}

\maketitle


\section{Introduction}
While General Relativity (GR) has been extraordinarily successful as the prevailing theory of gravity and has thus far passed every experimental test on local scales, it remains crucial to explore how we can probe new physics beyond GR. It is plausible that departures of GR may be significant on cosmological scales and at the high energies of the early universe.
And, Gravitational Waves (GWs) provide a new opportunity to test GR in such parameter range currently unexplored. This is because, at linear order, gravitational waves barely interact with matter and traverse the universe relatively unchanged. Thus, GWs  serve as an excellent probe for new physics at extremely high energies in the  early universe, such as modifications to GR. In fact, as far as we know, the universe may be filled with a cosmic GW Background (GWB), just like it is permeated with electromagnetic waves. Interestingly, several pulsar-timing array detectors have reported hints of an isotropic GWB \cite{NANOGrav:2023gor, Reardon:2023gzh, Xu:2023wog, EPTA:2023fyk, EPTA:2023gyr}, which could well have a primordial origin \cite{NANOGrav:2023hvm}.

One plausible component of the cosmic GWB are primordial GWs generated during inflation. For a review on GWs from inflation see Ref.~\cite{Guzzetti:2016mkm} and references therein. See also Ref.~\cite{Komatsu:2022nvu} for an overview of GWs and the polarization of the cosmic microwave background. However, the GW spectral density predicted from the simplest models of inflation has an amplitude too low to be detected by GW interferometers. Significantly, a resonant heavy field can amplify GW on the cosmic microwave background (CMB) scale while maintaining scale invariance in the curvature perturbation \cite{Cai:2021yvq}. Additionally, modifications to GR during inflation may give rise to a detectable blue-tilted GW spectrum, as investigated in relevant studies \cite{Kobayashi:2010cm, Endlich:2012pz, Cannone:2014uqa,Kawai:2023nqs}.
And, after inflation, modifications of GR could lead to a resonant amplification of GWs \cite{Lin:2015nda,Kuroyanagi:2017kfx,Cai:2020ovp,Nikandish:2023eak,Kawai:2017kqt}. Such resonances occur, e.g., if modifications of GR introduce oscillatory modulations in the linear GW equation.

In this paper, our aim is to carry out a general analysis on how modifications of GR may enhance a primordial GW signal due to the presence of narrow resonances. While there is a vast array of modified gravity theories, we focus on quadratic Degenerate Higher Order Scalar Tensor (DHOST) theories \cite{Langlois:2015cwa,Langlois:2015skt,Crisostomi:2016czh,BenAchour:2016cay}, which provide a general framework for scalar-tensor theories without the presence of the Ostrogradski ghost \cite{Woodard:2015zca,Ganz:2020skf}. 
In general, in this class of models the GW equation may be modified either directly by altering the friction term, the propagation speed of the GWs and/or indirectly by changing the background evolution due to the modified Friedmann equations. Note that all modifications depend on the background evolution of the scalar field.

A logical source of oscillatory modulations of the GW equations are background oscillations of the scalar field, e.g., if such scalar field is the dark matter. Then, the oscillations of the scalar field at the bottom of the potential, may lead to a parametric resonance resulting in an exponential enhancement of the GW strain. Since GWs travel cosmological distances, even minor periodic fluctuations can leave a significant imprints on the GW spectrum. For example, in Ref.~\cite{Cai:2020ovp} it has been discussed how oscillations in the propagation speed of GWs amplify a primordial wave to a detectable level by upcoming or current gravitational wave detectors. The concept of sound speed resonance (SSR) encompasses various phenomenological manifestations, as elucidated in prior works \cite{Cai:2018tuh, Cai:2019jah, Chen:2019zza, Chen:2020uhe}. Analogous investigations with a specific focus on induced oscillations atop the conventional Hubble evolution or modifications to the friction term have been conducted \cite{Ye:2023xyr, Nikandish:2023eak}. This form of parametric resonance effect finds application within the context of early cosmological scenarios, as explored in relevant studies \cite{Cai:2019bmk, Zhou:2020kkf, Pi:2021dft, Pi:2022zxs, Sasaki:2022rat}.

As a new approach to parametric resonances, we reduce the functional freedom of DHOST theories by use of disformal transformations \cite{Bekenstein:1992pj} to work in the “GW frame”, i.e. the DHOST frame in which the tensor modes propagate as in GR \cite{Creminelli:2014wna}. In this frame,  parametric resonances originating from the oscillations in the friction or propagation speed can all be associated to oscillations in the scale factor of the GW frame,  aiding us in identifying potential cancellation effects. For simplicity, we then take an Effective Field Theory (EFT) approach to DHOST and require that the modifications are subdominant, i.e. below the cut-off scale of the EFT. Focusing on the radiation dominated universe, we then show that: 
\begin{enumerate}
    \item[(1)] parametric resonances are suppressed in modified gravity theory that contain an Einstein frame, i.e., they can be mapped to GR via a metric transformation, and,
    \item[(2)] resonances can be sufficiently strong to enhance the primordial GW to detectable levels if there is no such Einstein frame. 
\end{enumerate}

We also apply our formalism to an ultra-light scalar field non-minimally coupled to gravity. Ultralight scalar fields has been extensively studied in the literature as a potential candidate for Ultralight Dark Matter (ULDM)
\cite{PhysRevD.28.1243,PhysRevLett.64.1084,Sin:1992bg,Hu:2000ke} (see Refs.~\cite{Ferreira:2020fam,Hui:2021tkt} for recent reviews). Recently, it has been shown that locally the ULDM can enhance the gravitational waves due to the oscillating gravitational potential \cite{Delgado:2023psl}. We demonstrate that in case of a simple non-minimal coupling, primordial gravitational waves can be significantly amplified on cosmological scales such that they could be detected by LISA, Einstein Telescope or Cosmic Explorer.

This paper is organized as follows. In \S~\ref{sec:DHOST}, we will recap the formulation of the DHOST theory and discuss how we can analyze these models in the GW frame by performing a disformal transformation. Further, we discuss the implications of the non-minimally coupled matter sector in the new gravitational wave frame. 
In \S~\ref{sec:Gravitational_wave_resonances_in_DHOST} we discuss how the parametric resonances in DHOST vary with the existence of an Einstein frame. Finally, in \S~\ref{sec:ULDM}, we apply our formalism to a specific example of an ultralight scalar field non-minimally coupled to gravity.

\section{DHOST theory and the GW frame}
\label{sec:DHOST}
We first briefly review the theory under study and then we present a convenient frame to study resonances in the GW equation. We work within the DHOST theory up to quadratic order as proposed in \cite{Langlois:2015cwa,BenAchour:2016cay}. The general action is given by
\begin{align}\label{eq:DHOSTaction}
	S = \int \mathrm{d}^{ 4 } x \sqrt{ - \tilde g } \Big[ & \tilde P( \phi, \tilde X ) + \tilde Q (\phi, \tilde X) \tilde \Box \phi + \tilde F ( \phi, \tilde X ) \tilde R + \sum_i a_i(\phi,\tilde X) \tilde L_i \Big] + S_m(\tilde g_{\mu\nu},\chi) ~ ,
\end{align}
where $\tilde X = - \tilde \nabla_\mu \phi \tilde \nabla^\mu \phi$ and $ \tilde L_i $ are defined as
\begin{align}
    & \tilde L_1 = \tilde \nabla _ { \mu } \tilde \nabla _ { \nu } \phi \tilde\nabla ^ { \mu } \tilde\nabla ^ { \nu } \phi ~ , \qquad
    \tilde L_2= ( \tilde \Box \phi ) ^ 2 ~ , \qquad \tilde L_3 = ( \tilde \Box \phi ) \tilde\nabla ^ { \mu } \phi \tilde\nabla _ { \mu } \tilde\nabla _ { \nu } \phi \tilde\nabla ^ { \nu } \phi ~ , \\
    &\tilde L_4 = \tilde\nabla _ { \mu } \tilde\nabla _ { \nu } \phi \tilde\nabla ^ { \mu } \phi \tilde\nabla ^ { \nu } \tilde\nabla _ { \rho } \phi \tilde\nabla ^ { \rho } \phi ~ , \qquad \tilde L_5= ( \tilde\nabla _ { \mu } \tilde\nabla _ { \nu } \phi \tilde\nabla ^ { \mu } \tilde\nabla ^ { \nu } \phi ) ^ 2 ~ .
\end{align}
Note that out of the six free functions $\tilde F$ and $\tilde a_i$, only three are independent. The others are constrained by the DHOST degeneracy conditions. In the following we focus on Class I of the quadratic DHOST models \cite{BenAchour:2016cay}, which includes Horndeski and GR as a subclass and is not plagued by pathologies \cite{Langlois:2017mxy}, where
\begin{align}\label{eq:conditionsclassI}
   \tilde a_1 =& -\tilde a_2 ~ , \\
   \tilde a_4 =& \frac{1}{8 (\tilde F- \tilde X \tilde a_2)^2} \Big[ -16 \tilde X^2 \tilde a_2^3 + 4 (3\tilde F+ 16 \tilde F_{\tilde X} \tilde X) \tilde a_2^2 + (-16 \tilde X^2 \tilde F_{\tilde X} + 12 \tilde X \tilde F) \tilde a_2 \tilde a_3 - \tilde X^2 \tilde F \tilde a_3^2 \nonumber \\
    & - 16 \tilde F_{\tilde X} (3\tilde F + 4 \tilde F_{\tilde X} \tilde X) \tilde a_2 + 8 \tilde F (\tilde X \tilde F_{\tilde X} - \tilde F) \tilde a_3 + 48 \tilde F \tilde F_{\tilde X}^2  \Big] ~ ,\\
    \tilde a_5=& \frac{(-4 \tilde F_{\tilde X}+ 2 \tilde a_2 - \tilde X \tilde a_3) (-2 \tilde a_2^2 - 3 \tilde X \tilde a_2 \tilde a_3 + 4 \tilde F_{\tilde X} \tilde a_2 + 4 \tilde F \tilde a_3)}{8 (\tilde F - \tilde X \tilde a_2)^2} ~ .
\end{align}

We then consider the perturbed Friedmann–Lemaître–Robertson–Walker (FLRW) metric given by
\begin{align}\label{eq:FLRWmetric}
    \mathrm{d} \tilde s^2 = - \tilde a^2 \mathrm{d} \tau^2 + \tilde a^2 (e^{\tilde h})_{ij} \mathrm{d} x^i \mathrm{d} x^j \,, 
\end{align}
where ${\tilde h}_{ij}$ are the tensor perturbations, or alternatively the transverse-traceless degrees of freedom. Note that for simplicity we only focus on the tensor perturbations. The quadratic action for tensor perturbations reads
\begin{equation}
	S _ { T } ^ { ( 2 ) } = \int \mathrm{d}^{3}x d\tau \tilde a ^ 2 \frac{ \tilde M ^ 2 }{ 2 } \left[ \tilde h^{\prime 2} _ { i j } - ( 1 + \tilde \alpha _ { \rm T } ) ( \partial \tilde h _ { i j } ) ^ 2 \right] ~ , 
\end{equation}
where
\begin{equation}
	\frac{ \tilde M ^ 2 }{ 2 } =\tilde F - \tilde a_{ 1 } \tilde X ~ , \qquad \frac{ \tilde M ^ 2 }{ 2 } ( 1 + \tilde \alpha _ { \rm T } ) = \tilde F ~ .
\end{equation}
From now on we work in Fourier space, where in our convention we have
\begin{align}
    \tilde h_{ij}(\mathbf{x},\tau) = \int \mathrm{d}^3k e^{i \mathbf{k} \mathbf{x}} \tilde h_k^\lambda(\tau) \epsilon^\lambda_{ij}(\hat{\mathbf{k}}) ~ \,.
\end{align}
Then, introducing the canonical normalized variable $\tilde v = \tilde a \tilde M \tilde h^\lambda_k$, the equation of motion for both of the tensor modes is given by
\begin{align}\label{eq:canonicaltensor}
    \tilde v^{\prime\prime} + \left( (1+\tilde \alpha_T) k^2 - \frac{(\tilde a\tilde M)^{\prime\prime}}{\tilde a \tilde M} \right) \tilde v=0 ~ .
\end{align}

From Eq.~\eqref{eq:canonicaltensor} we see that, general speaking, modified gravity influences the evolution of the tensor modes in two distinct ways. Firstly, it can alter the sound speed via $\tilde\alpha_T$ and the mass term via $\tilde M$ (which is equivalent to the friction for the non-canonical normalized tensor modes). Secondly, it affects the evolution of the Hubble parameter/scale factor $\tilde a$ via the modified Friedmann equation. However, as is the case of a simple non-minimal coupling like $f(\phi)\tilde R$, the effects from $\tilde a$ and $\tilde M$ may cancel each other (see, e.g., App.~\ref{app:Conformal_matter}). We find that to simplify the discussion, it is more convenient to perform the calculations in what we call the “GW frame”, where the tensor modes propagate as in GR, i.e. $M=0$ and $\alpha_T=0$. Note that hereon we drop the tildes for quantities in the GW frame. In that frame, all the modified gravity effects are solely encoded in the modifications of the scale factor. Note that a similar approach has been used in \cite{Creminelli:2014wna,Fujita:2015ymn} to show that inflationary predictions of the tensor power spectrum are directly given by the quantities in the GW frame.

\subsection{The GW frame}
\label{sec:Gravitational_wave_frame}
As we shall see, it is convenient to rewrite the starting DHOST theory to its GW frame by employing a general disformal transformation given by
\begin{equation}\label{eq:disformaltransformation}
	\tilde g _ { \mu \nu } = C ( \phi, X ) g _ { \mu \nu } + D ( \phi, X ) \nabla _ { \mu } \phi \nabla _ { \nu } \phi ~\,,
\end{equation}
where $C$ and $D$ are functions fixed by the condition of the GW frame, as we shall shortly see. We also impose the condition that $C-D X>0$ to have a well-defined signature of the metric $\tilde g _ { \mu \nu }$.
The functions $\tilde F$ and $\tilde a_1$ in Eq.~\eqref{eq:DHOSTaction} are related to $F$ and $a_1$ after the transformation \eqref{eq:disformaltransformation} via the formulas provided in Refs.~\cite{BenAchour:2016cay,Langlois:2017mxy}, namely
\begin{align}\label{eq:relations}
    F = \sqrt{C} \sqrt{C-D X} \tilde F ~ , \qquad a_1 = C^{3/2} \sqrt{C-D X} \left( \frac{D}{C(C-D X)} \tilde F + \frac{1}{(C - D X)^2} \tilde a_1 \right) ~ \,.
\end{align}
By definition, the GW frame has $  M^2= 1 $ and $  \alpha _ { \rm T } = 0 $, which implies
\begin{equation}\label{eq:conditionsGWframe}
	  F = \frac{  M ^ 2 }{ 2 } ( 1 +  \alpha _ { \rm T } ) ~ , \quad  a _ { 1 } = \frac{  M ^ 2 }{ 2 } \frac{  \alpha _ { \rm T } }{ X } \quad \Rightarrow \quad  F = \frac{ 1 }{ 2 } ~ , \quad  a _ { 1 } = 0 ~ .
\end{equation}
This also means that in the GW frame the equations of motion for the canonical tensor modes are given by 
\begin{align}\label{eq:GWequationframe}
    v^{\prime\prime} + \left( k^2 - \frac{a^{\prime\prime}}{a} \right) v =0 ~ .
\end{align}
As is apparent, all the modifications must be encoded in the scale factor.

As we anticipated, the GW frame condition Eq.~\eqref{eq:conditionsGWframe} fixes the functions in the disformal transformation \eqref{eq:disformaltransformation}. After solving Eq.~\eqref{eq:relations} using Eq.~\eqref{eq:conditionsGWframe} we find that
\begin{align}
\label{eq:disformal_relations}
    \hat D= - 4 \tilde F \tilde a_1  C ~ , \qquad 
    C^3 + \frac{1}{4 \tilde F \tilde a_1 X} C^2 - \frac{1}{16 \tilde F^3 \tilde a_1 X} =0 ~ ,
\end{align}
where for later convenience we introduced $\hat D = D/C$. Note that $\tilde F$ should be understood as $\tilde F=\tilde F(\phi,X/(C(1-\hat D X)))$. With Eq.~\eqref{eq:disformal_relations}, we obtain that the general action in the GW frame is given by
\begin{align}
\label{eq:Action_Gravitational_wave_frame}
    S = \int \mathrm{d}^4x\, \sqrt{-g} \Big[ \frac{1}{2} R + P + Q \Box \phi + a_3 L_3 + a_4 L_4 + a_5 L_5 \Big] + S_M(\tilde g_{\mu\nu},\chi) ~ ,
\end{align}
where note that matter is coupled to the original metric $\tilde g_{\mu\nu}$. Also note that the classes of DHOST are invariant under disformal transformations and, therefore, using the condition for Class I \eqref{eq:conditionsclassI} we obtain $a_1=a_2=0$. In the GW frame the original functions $\tilde F$, $\tilde a_2$ and $\tilde a_3$ are replaced by $a_3$, $C$ and $D$. The specific form of the functions $a_i$, $Q$ and $P$ in terms of the old functions are not relevant for our purpose and we refer the interested reader to Ref.~\cite{BenAchour:2016cay}. 
The degeneracy conditions for $a_4$ and $a_5$ simplify to
\begin{align}
    a_4 = - \frac{1}{4} \left( 4a_3 + a_3^2 X^2 \right) ~ , \qquad 
    a_5= -a_3^2 X ~ .
\end{align}

At this point it is important to clarify the meaning and usefulness of the disformal transformation \eqref{eq:disformaltransformation}. This is clear if we expand the perturbed FLRW metric \eqref{eq:FLRWmetric} in conformal coordinates and compare the scale factors and the tensor perturbations at leading order, which yields
\begin{align}\label{eq:disformalscale}
\tilde a &= C^{1/2} a \,,\\
    \tilde h _ {ij} &=  h _ { ij } + \hat D ( \phi, X )\left(\partial _ { i } \delta\phi \partial _ { j } \delta\phi \right)_{\rm TT}~\,,\label{eq:disformaltensor}
\end{align}
where $TT$ refers to the transverse-traceless component. Note that we considered perturbations of the scalar field, namely $\phi\to\phi+\delta\phi$, to illustrate in Eq.~\eqref{eq:disformaltensor} that the effects of the function $\hat D$ to $\tilde h_{ij}$ only appear at second order in perturbation theory.\footnote{This effect might be interesting for secondary  GWs (see, e.g., Ref.~\cite{Domenech:2021ztg} for a review) but it is out of the scope of this paper. We will investigate this possibility in future works.} Thus, we see that leading order $\tilde h_{ij}=h_{ij}$. This is also consistent with the fact that GW luminosity distance depends on the scale factor of the GW frame \cite{Belgacem:2017ihm,LISACosmologyWorkingGroup:2019mwx}, since on subhorizon scales we have that $\tilde h_{ij}=h_{ij}\propto 1/a$.

One may wonder whether fast oscillatory modulations of the scale factor would also appear in GW detectors, since matter fields would effectively see the combination $\tilde a^2(\delta_{ij}+\tilde h_{ij})$, e.g., that appears in the volume element. However, the oscillatory modulations must yield small corrections to the expansion rate and are therefore negligible for local measurements. In any case, we eventually require that the effects of modified gravity are only present at  early times and that these modifications vanish today. Namely we impose $C(t_0)=1$ and $D(t_0)=0$, $t_0$ refers to today. Consequently, the two frames (i.e. the matter frame and the GW frame) coincide at present. Thus, in summary, resonant amplifications of $\tilde h_{ij}$ can be fully understood by studying resonant amplifications of $h_{ij}$.

\subsection{Matter fields in the GW frame}
\label{sec:Matter_coupling}
For completeness we study the equations for matter fields in the GW frame before moving on to the GW resonances. As an example for the matter sector, we model radiation as a k-essence fluid whose action is given by
\begin{align}
    S_m = \int \mathrm{d}^4x\, \sqrt{-\tilde g}  \left( - \frac{1}{2} \tilde g^{\mu\nu} \partial_\mu \chi \partial_\nu \chi \right)^2 \equiv \int \mathrm{d}^4x\,\sqrt{-g} \tilde Y^2 ~ .
\end{align}
Working in variables of the GW frame we have that
\begin{align}
    S_m = \int \mathrm{d}^4x\,\sqrt{- g}  \sqrt{1- \hat D X}  \left( Y + \frac{1}{2} \frac{\hat D}{1-\hat D X} (\partial^\mu \phi \partial_\mu \chi)^2 \right)^2 ~ .
\end{align}
Note that in the GW frame, matter depends only on $\hat D$ since radiation is conformal invariant. 
The equation of motion for the matter field $\chi$ can be written as
\begin{align}
    \nabla_\alpha J^\alpha =0 ~ .
\end{align}

Since we later take an EFT expansion, let us assume that the leading order contribution to $\hat D$ is independent of $X$, namely $\hat D=d_0(\phi)/\Lambda^2$. Then, at the background level, we obtain for radiation  at leading order
\begin{align}
    J^\alpha \simeq - \delta^{\alpha}_0 \dot \chi^3 \left(  1 -\frac{3\dot \phi^2}{2 \Lambda^2}   d_0  \right) ~ ,
\end{align}
which yields
\begin{align}
     \dot \rho_r +  3 H (\rho_r + p_r)  =& \left(1- \frac{3d_0 \dot \phi^2}{2\Lambda^2} \right)^{-1}  12 p_r \left( \frac{ d_0 \dot \phi \ddot \phi}{\Lambda^2} + \dot d_0 \frac{\dot \phi^2 }{2 \Lambda^2} \right) \nonumber \\
      \simeq &   p_r \left(12 \frac{ d_0 \dot \phi \ddot \phi}{\Lambda^2} + 6 \dot d_0 \frac{\dot \phi^2 }{\Lambda^2}  \right) ~ ,
\end{align}
where $p_r$ and $\rho_r$ are the pressure and energy density of radiation respectively, and $p_r=\frac{\rho_r}{3}=\frac{  \dot \chi^4}{4}$ .

The energy momentum tensor $T^{m}_{\mu\nu}$ at the FLRW background can be expressed as
\begin{align}
    T^m_{\alpha\beta}=& \delta_\alpha^0 \delta^0_\beta \left( \rho_r +  \frac{11}{2} d_0 p_r \frac{\dot \phi^2}{\Lambda^2}  \right) + \delta_{\alpha i} \delta_{\beta j} a^2 p_r \left( 1 + \frac{3}{2} \frac{\dot \phi^2}{\Lambda^2} \right) ~ .
\end{align}
Later on, we will show that the coupling of the scalar field to the matter sector is suppressed and will not lead to sizeable resonances.

\section{GW resonances in DHOST}
\label{sec:Gravitational_wave_resonances_in_DHOST}
Let us  explore the general possibilities for parametric resonances in the GW equation in DHOST. As a general setup, and for analytical viability, we assume that the background evolution is dominated by radiation and the modifications stemming from DHOST are subdominant. We start with the simplest case of minimally coupled scalar fields and show that resonances are in general suppressed. We then investigate two different classes of models of DHOST: $(1)$ those with an Einstein frame (that is $a_i = Q =0$) and (2) those without (i.e. $a_i \neq 0$ and $Q \neq 0$). We find that in the former resonances are suppressed, as in the minimally coupled case, while, in the latter, there may be sizable resonances in the GW strain.
 
From now on and for simplicity, we take an EFT-inspired approach and assume that we one can expand the free functions in powers of $X/\Lambda^2$, where $\Lambda$ could be identified with the cutoff scale of our theory.\footnote{Strictly speaking, we consider DHOST to be the full theory, without higher order corrections. Then, for analytical viability, we assume that modifications of gravity are small corrections to GR and can, therefore, be treated in a perturbative manner. We choose $\dot\phi/\Lambda$ as the expansion parameter of such corrections because we are interested in the kinetic dependence of the non-minimal couplings, which is responsible, e.g., for modifications in $c_T$.} In this case, the leading order terms scale as
\begin{align}
    & C = c_0(\phi) + c_1(\phi) \frac{X}{\Lambda^2} ~ , \qquad \hat D = \frac{d_0(\phi)}{\Lambda^2} ~ , \qquad a_3= \frac{a_{3,0}(\phi)}{\Lambda^4} ~ , \nonumber \\
    & P = \frac{1}{2} P_0(\phi) X - V(\phi) + P_1(\phi) \frac{X^2}{\Lambda^2} ~ , \qquad Q = Q_1(\phi) \frac{X}{\Lambda^2} ~ .
    \label{eq:Form_of_Functions}
\end{align}
Note that we can always redefine the scalar field to set $P_0(\phi)=1$. In passing, let us clarify that the main goal of our paper is to provide a proof of concept on the existence of parametric resonances due to modifications of gravity. The purpose of the expansion in $X/\Lambda^2$ is to identify under which conditions a non-minimal coupling can lead to parametric resonances. The chosen EFT-inspired expansion is the most simplistic ansatz which is consistent with our assumption that we recover GR to a very good approximation in the late universe. It is important to note though that, in general, there is no guarantee that the DHOST constraint structure is protected from quantum corrections. This is not necessarily a problem if one suitable relaxed the DHOST structure to begin with--a strategy which has been called the "scordatura mechanism" \cite{Motohashi:2019ymr}. Nevertheless, in section \ref{sec:ULDM}, we will apply our analysis to a specific model in the Horndeski subclass which is well defined as a formal EFT expansion if there is a weakly broken Galilean symmetry \cite{Pirtskhalava:2015nla,Santoni:2018rrx}.

\subsection{Parametric resonances for minimally coupled fields}
\label{subsec:Simple_parametric_resonance}
To gain an understanding of potential parametric resonances, we begin with the simplest case of a minimally coupled canonical scalar field oscillating around the bottom of the potential, which corresponds to  $ a_i = Q=P_1=C = \hat D = 0$ and $P_0(\phi)=1$ in Eqs.~\eqref{eq:Action_Gravitational_wave_frame} and \eqref{eq:Form_of_Functions}. In simpler terms, the GW frame coincides with the matter frame, which is GR plus a canonical scalar field. This leads to minor oscillations on top of the standard evolution of the Hubble parameter. This scenario has  been extensively examined in existing literature, as seen in \cite{Alsarraj:2021yve,Ye:2023xyr,Nikandish:2023eak}, hence we will just concentrate on the key points.

Given the scalar field oscillates at the bottom of the potential, we parameterize it as
\begin{align}
  \phi = \bar \phi(\tau) f(\omega \tau)~,   
\end{align}
where $\bar \phi$ is the slowly varying amplitude and $\vert f(\omega \tau)\vert \leq 1$ corresponds to rapid oscillations. Additionally, we assume that $\omega \gg \mathcal{H}$, with $\mathcal{H}= a^\prime/ a$ being the conformal Hubble parameter. Utilizing this, we can approximate the equation of motion as
\begin{align}
\label{eq:simple_parametric_resonance}
    \frac{\mathrm{d}^2\tilde v}{\mathrm{d}x^2} + \left( k^2 + \frac{\bar \phi^2 }{6 \tilde a^2} f^{\prime 2}  - \frac{2\tilde V}{3 \omega^2} \right) \tilde v =0 ~ ,
\end{align}
where $x = \omega \tau$. For instance, for a quartic potential $V(\phi) = \lambda \phi^4$, the scalar field frequency is given by $\omega = \sqrt{\lambda \bar \phi^2 a^2/2}$ and $f(\omega \tau) = \mathrm{sn}(\omega \tau,-1)$, with $\mathrm{sn}$ being the Jacobian elliptic function (refer to section \ref{sec:ULDM} for more details). For $\omega \gg \mathcal{H}$, we can neglect the time dependence of $\bar \phi$ and $a$ compared to the rapid oscillation in $f^\prime(x)$ or $f(x)$, resulting in a Hill's equation. The parametric resonance is, however, suppressed by the amplitude of the scalar field $\bar \phi^2$. It is important to note that we are working in units where the Planck mass is set to one. Therefore, in this case the resonance is large only if the the scalar field amplitude is near to the Planck scale, otherwise it may yield subdominant modulations to the GW spectrum.

In what follows, we use the oscillating, minimally coupled, canonical scalar field as a reference to check whether resonances can be enhanced by modified gravity or not with respect to such benchmark.

\subsection{Resonances in DHOST models with Einstein Frame}
\label{sec:Einstein_frame}
As a next step, let us consider the case where the modified gravity model has an Einstein frame, i.e., after performing the disformal transformation we have that $a_i=0$ and $Q=0$. Recall that, however, the matter sector is disformally coupled to the scalar field. The total action then reads
\begin{align}
    S = \int \mathrm{d}^4x\, \sqrt{-g} \Big[ \frac{1}{2} R + P(\phi,X) \Big] + S_M(\tilde g_{\mu\nu},\chi) ~ .
\end{align}
In this case, the EOM takes the schematic form:
\begin{align}
    G_{\alpha\beta} = T^P_{\alpha\beta} + T^m_{\alpha\beta} ~ ,
\end{align}
where $T^P_{\alpha\beta}$ is the energy momentum tensors of the scalar field. At the FLRW background, it is given by
\begin{align}
    T^P_{\alpha\beta} = \delta_{\alpha}^0 \delta_{\beta}^0 \rho_\phi + \delta_{\alpha i} \delta_{\beta j} a^2 p_\phi = \delta_{\alpha}^0 \delta_{\beta}^0 \left( \frac{1}{2} X + V + 3 P_1 \frac{X^2}{\Lambda} \right) + \delta_{\alpha i} \delta_{\beta j} a^2 \left( \frac{1}{2} X - V + P_1 \frac{X^2}{\Lambda^2} \right)~,
\end{align}
where in the second step we have used the expansion of $P(\phi,X)$ given by Eq.~\eqref{eq:Form_of_Functions}.
Using the background equation of motions in conformal time we obtain
\begin{align}
    -2 \frac{a^{\prime\prime}}{a} =  \left( p_\phi - \frac{1}{3}\rho_\phi\right) a^2  - \frac{1}{3} d_0 p_r a^2 \frac{\dot \phi^2}{\Lambda^2} ~ ,
\end{align}
resulting in
\begin{align}
    v^{\prime\prime} + \left( k^2 + \frac{1}{2} \left(p_\phi - \frac{1}{3} \rho_\phi \right) a^2 - \frac{1}{6} d_0 p a^2 \frac{\dot \phi^2}{\Lambda^2} \right) v\simeq 0 ~ .
\end{align}
Firstly, note that if the original model is related to GR via a pure conformal transformation (i.e. $D=0$), such as the Brans-Dicke model studied in App.~\ref{app:Conformal_matter}, the last term vanishes and there is no direct dependence on the matter sector due to the conformal invariance of radiation. The structure of the equation of motion is essentially the same as in \eqref{eq:simple_parametric_resonance} and the resonance is highly suppressed for $\bar \phi \ll 1$. 

Now we consider two different regimes: deep in radiation domination where $\rho_\phi \ll p_r \dot \phi^2/\Lambda^2$, and the scalar field domination where $p_r  \dot \phi^2/\Lambda^2 \ll \rho_\phi $. 

\subsubsection{Radiation domination}
In the deep radiation domination epoch, we can approximate $p_r \dot \phi^2/\Lambda^2 \gg \rho_\phi, p_\phi$, rendering only the last term significant. Given that term is already suppressed by $\dot \phi^2/\Lambda^2$, we only consider the leading order solution of $\phi$ and $p_r$. As before, we model the background solution of $\phi$ as $\phi = \bar \phi(\tau) f(\omega \tau)$. Then, the EOM can be approximated as
\begin{align}
    \frac{\mathrm{d}v}{\mathrm{d}x^2} + \left( \frac{k^2}{\omega^2} - \frac{1}{6} d_0 \frac{\mathcal{H}^2 \bar \phi^2}{a^2 \Lambda^2} f^{\prime 2}(x)  \right) v\simeq 0 ~ ,
\end{align}
where $x=\omega \tau$ and we have used that $p_r a^2 \simeq \mathcal{H}^2$ from the background EOM. The equation of motion again takes the form of a Hill's equation. However, the parametric resonance is suppressed by $\frac{\mathcal{H}^2 \bar \phi^2}{a^2 \Lambda^2}$. To maintain a consistent expansion in orders of $X/\Lambda^2$, we must require that $X/\Lambda^2 =\bar \phi^2 \omega^2/(a^2 \Lambda^2) \ll 1$. Therefore, at best, the resonance is suppressed by $\mathcal{H}^2/\omega^2$. However, to neglect the time dependency of $\mathcal{H}$ and $\bar \phi$, we need to require that $\omega \gg \mathcal{H}$, which shows that the resonance is highly suppressed. 

\subsubsection{Scalar field domination}
The other possibility is that the energy density of the scalar field is not subdominant, so that we cannot neglect $p_\phi-\rho_\phi/3$. Using the form of $P(\phi,X)$ \eqref{eq:Form_of_Functions}, we obtain
\begin{align}
    p_\phi - \frac{1}{3}\rho_\phi = \frac{1}{3} X - \frac{4}{3} V ~ ,
\end{align}
which takes the same form as for the canonical scalar field in \eqref{eq:simple_parametric_resonance}, since the leading contribution proportional to $P_1(\phi)$ cancels each other. 
Therefore, the parametric resonance is suppressed by $\bar \phi^2$ (see discussion in the previous section \ref{subsec:Simple_parametric_resonance}). 

In summary, as long as the modified gravity model can be expressed in terms of a K-essence fluid, whose scalar amplitude is much smaller than the Planck scale, and a non-minimally coupled radiation fluid, the resonances are highly suppressed. Note that this might change if we consider a different matter sector which is not conformal invariant (see for instance \cite{Nikandish:2023eak}). 

\subsection{Resonances in DHOST models with No Einstein Frame}
\label{sec:No_Einstein_frame}
As a final step, we consider the general case where it is not possible to eliminate all higher derivative terms in the action via the disformal transformation. The Einstein Equations can now be schematically written as
\begin{align}
    G_{\alpha\beta} = T^P_{\alpha\beta} + T^m_{\alpha\beta} + T^{Q}_{\alpha\beta} + T^{a}_{\alpha\beta} ~ ,
\end{align}
where $T^{Q}_{\alpha\beta}$ and $T^a_{\alpha\beta}$ denote the contributions from the non-minimally coupled scalar field.
At the background level, they are explicitly given by
\begin{align}
    T^Q_{\alpha\beta} =& \delta_\alpha^0 \delta_\beta^0 \left( - 6 H Q_1 \frac{\dot \phi^3}{\Lambda^2} + Q_1^\prime \frac{\dot \phi^4}{\Lambda^2} \right) + \delta_{\alpha i} \delta_{\beta i} a^2 \left( 2 Q_1 \frac{\dot \phi^2 \ddot \phi}{\Lambda^2} + Q_1^\prime \frac{\dot \phi^4}{\Lambda^2} \right) ~ , \\
    T^a_{\alpha\beta}=& \delta_\alpha^0 \delta_\beta^0 \left( a_{0} \left( (9 H^2 + 3 \dot H) \frac{\dot \phi^4}{\Lambda^4} - 3 H \frac{\dot \phi^3 \ddot \phi}{\Lambda^4} \right) + 3 a_{0}^\prime H \frac{\dot \phi^5}{\Lambda^4} \right) \nonumber \\
    & + \delta_{\alpha i} \delta_{\beta i} a^2 \left( a_{0} \frac{3 \dot \phi^2 \ddot \phi^2 + \dot \phi^3 \dddot \phi }{\Lambda^4} + a_{0}^\prime \frac{\dot \phi^4 \ddot \phi}{\Lambda^4} \right) ~ .
\end{align}
Assuming that $\omega \gg H$, $\phi \ll 1$ and using the ansatz $\phi=\bar \phi(\tau) f(\omega \tau)$ we obtain
\begin{align}
    \frac{\mathrm{d}^2 v}{\mathrm{d}x^2} + \left( \frac{k^2}{\omega^2} + Q_1 \frac{ \omega^2 \bar \phi^3}{a^2 \Lambda^2} f^{\prime 2} f^{\prime\prime}  + \frac{1}{2} a_{0} \frac{\bar \phi^4 \omega^4}{a^4 \Lambda^4} (3 f^{\prime 2} ( f^{\prime\prime})^2 + f^{\prime 3} f^{\prime\prime\prime}) \right) v \simeq 0 ~ .
\end{align}
We see that the leading order terms relevant for the resonances are suppressed by $ Q_1 X/\Lambda^2 \bar \phi \sim Q_1 \omega^2 \bar \phi^2/(a^2\Lambda^2) \bar \phi$ or $ a_{0} X^2/\Lambda^4 \sim \omega^4 \bar \phi^4/(a^4\Lambda^4)$. These terms are dominant in comparison to a standard canonical scalar field if $X/\Lambda^2 \sim \bar \phi^2 \omega^2/(a^2\Lambda^2) \geq \bar \phi$. 

Therefore, for a very small amplitude of the scalar field, the dominant contributions will come from $a_i$ and then $Q$, but they are suppressed by $X^2/\Lambda^4$ or $(X/\Lambda^2) \phi$, instead of just $X/\Lambda^2$ as one might expect. This is interesting since in the original frame, the modifications of the sound speed or the friction term scale at leading order as $\tilde X/\Lambda^2$. However, as we can see from the gravitational wave frame, these modifications each other at leading order so that the final resonance is suppressed by $\tilde X^2/\Lambda^4$ or $\phi \tilde X/\Lambda^2$.

\section{Non-minimally coupled ultra-light dark matter}
\label{sec:ULDM}
Let us turn to an interesting application of our formalism. We consider ULDM with a non-minimal coupling to gravity, where the total action reads
\begin{align}\label{eq:uldmaction}
    S = \int \mathrm{d}^4\,x \sqrt{-\tilde g} \Big[& \frac{1}{2} \tilde R  + \frac{1}{2} \tilde X - \frac{1}{2} m^2 \phi^2 - \frac{\lambda}{4} \phi^4 +\tilde G^{\mu\nu} \frac{\phi_\mu \phi_\nu}{\Lambda^2} \Big] + S_m(\tilde g_{\mu\nu}, \psi) ~ ,
\end{align}
Interestingly, this specific form of the non-minimal coupling has been discussed in the literature in the context of galaxy clustering \cite{Gandolfi:2023hwx}.\footnote{We can also rewrite the action into a more standard form after integration by parts, namely
\begin{align}
     S = \int \mathrm{d}^4\,x \sqrt{-\tilde g} \Big[& \frac{1}{2} \left( 1+ \frac{2 \tilde X}{\Lambda^2}\right) \tilde R  + \frac{1}{2} \tilde X - \frac{1}{2} m^2 \phi^2 - \frac{\lambda}{4} \phi^4 - \frac{1}{\Lambda^2} \left( \tilde \nabla_\mu \tilde\nabla_\nu \phi \tilde\nabla^\mu \tilde \nabla^\nu \phi - (\tilde \Box \phi)^2 \right) \Big] + S_m(\tilde g_{\mu\nu}, \psi) ~ .
\end{align}} 
This model belongs to the largely studied Horndeski theory (see, e.g., Ref.~\cite{Kobayashi:2019hrl} for a review). In particular, we have checked that the model given by the action \eqref{eq:uldmaction} fulfills all the stability conditions at the linear level around the FLRW background as long as $\tilde X/\Lambda^2 \ll 1$. Note that, although one might worry about the possibility of a superluminal propagation speed of GWs, i.e. $\tilde c_T^2>1$, it does not necessarily indicate a violation of causality, as argued in Ref.~\cite{deRham:2020zyh}.  Nevertheless, let us emphasize that, as mentioned in the introduction, we consider only those models in which modifications to $c_T$ are relevant in the early universe and become negligible today.  Most interestingly, we find that, within the allowed mass range of ULDM, namely from $10^{-24}\,\mathrm{eV}$ to $1\,\mathrm{eV}$ (see, e.g., Ref.~\cite{Ferreira:2020fam} for a review), resonances due to modified gravity can enhance the amplitude of the GW background in the range of current or future GW detectors. We will show this in detail later.  

As we argued in \S~\ref{sec:Gravitational_wave_resonances_in_DHOST}, it is more convenient to work in the GW frame. In the present case, the GW frame is obtained from Eq.~\eqref{eq:uldmaction} after performing a disformal transformation \eqref{eq:disformaltransformation} with
\begin{align}\label{eq:CCCC}
    C =& - \frac{1+ s - \sqrt{2} \frac{X}{\Lambda^2} c }{8 \frac{X}{\Lambda^2}} \simeq 1 - \frac{6 X}{\Lambda^2} ~ , \\
    \hat D=& \frac{1}{\Lambda^2} \frac{8 + \sqrt{2} (-1 + s) c}{16 \frac{X}{\Lambda^2}} \simeq - \frac{2}{\Lambda^2} + \frac{8 X}{\Lambda^4} ~ ,\label{eq:DDDD}
\end{align}
where we defined
\begin{align}
    c = \sqrt{\frac{1+ s+ 4 \frac{X}{\Lambda^2} \left( 1+ 2 s - 4 (3+s) \frac{X}{\Lambda^2} \right) }{\frac{X^2}{\Lambda^4}}} ~ , \qquad s = \sqrt{1 - 8 \frac{X}{\Lambda^2}} ~\,,
\end{align}
and in the last step of Eqs.~\eqref{eq:CCCC} and \eqref{eq:DDDD} we expanded at leading order in $X/\Lambda^2$. After the disformal transformation we find in the GW frame that
\begin{align}
    P(\phi,X) =& \sqrt{1- \hat D X} C^2 \left[ \frac{1}{2} \frac{X}{C(1-\hat D X)} - V(\phi) \right] \simeq \frac{1}{2} X - V + \frac{X}{\Lambda^2} \left( 11 V - \frac{7}{2} X\right), \\
    a_3(\phi,X) \simeq & \frac{1}{\Lambda^4} \left( 40 + 328 \frac{X}{\Lambda^2} \right) ~ .
\end{align}
We have $Q=0$ since except for the potential, the free functions are shift symmetric. 

Since we are interested in small oscillatory modulations, we may only consider the leading order solution of the background equation of motion for $\phi$, which is given by
\begin{align}
    \ddot \phi + 3 H \dot \phi + m^2 \phi + \lambda \phi^3 =& \mathcal{O}(\frac{X}{\Lambda^2}) ~ .
\end{align}
In the regime where the self-interaction term dominates $\lambda \phi^2 \gg m^2$, which we expect to be the case in the  early universe for light scalar fields, we obtain
\begin{align}
    \phi \simeq \frac{\phi_0}{a}\, \mathrm{sn}(x,-1) \qquad,\qquad X \simeq \frac{\phi_0^2 \omega^2 }{2 a^4} \left( \mathrm{cn}(x,-1) \mathrm{dn}(x,-1) - \frac{\mathcal{H}}{\omega} \mathrm{sn}(x,-1) \right)^2 ~,
\end{align}
where
\begin{align}
    x = \omega (\tau - \tau_\star)~, \qquad \omega = \sqrt{\lambda \phi_0^2/2}~,
\end{align}
and $\mathrm{sn}(x,-1)$, $\mathrm{dn}(x,-1)$ and $\mathrm{cn}(x,-1)$ are  the Jacobi elliptic functions and $\tau_\star$ is the conformal time at the start of the resonance period. As an abuse of notation, we will use $\mathrm{sn}(x,-1)\equiv \mathrm{sn}(x)$ and similarly for $\mathrm{cn}(x)$ and $\mathrm{dn}(x)$.
In terms of our previous notation this solution corresponds to  $\bar \phi(\tau) = \phi_0/a$ and $f(\omega \tau)= \mathrm{sn}(x)$.

One may worry that self-interactions of the scalar field may backreact or change the oscillatory behavior and spoil the resonance. In order to estimate the effect, consider the linear scalar perturbations, say $\delta\phi$, around the background equation of motion. By expanding in orders of $X/\Lambda^2$ and using that $\omega \gg \mathcal{H}$, we have checked that self-interactions will be dominated by the quartic interaction, yielding an effective mass to $\delta\phi$ proportional to the standard $\lambda \phi^2$. In the end, this amounts to order one (normalized to the resonance frequency) oscillations. The other terms coming from modified gravity have a similar form but are suppressed by factors of $X/\Lambda^2$. It is well known that the quartic self-interactions do not spoil the oscillatory behavior/resonance (see for instance \cite{Figueroa:2016wxr} for a lattice simulation of the quartic interaction in the context of reheating) and, therefore, we expect that this also applies to the present case as long as the  modified gravity terms are subdominant.

Note that when the quartic self-interaction dominates, the scalar field behaves as radiation. Only after the mass term dominates, which happens for $\tau \geq \tau_t$, where $\tau_t$ is defined through $m^2 a(\tau_t)^2 = \lambda \phi_0^2$, the scalar field behaves as dark matter, i.e. $\bar \phi \propto 1/a^{3/2}$. We use the subscript “$t$” in $\tau_t$ to denote the transition to the time when $m^2\phi^2$ dominates the potential. Of course, we require this to occur before matter-radiation equality, namely $\tau_t < \tau_{eq}$. Requiring that the scalar field is responsible for all dark matter at radiation-matter equality we find the amplitude $\phi_0$ to be roughly $\phi_0 \sim \frac{\mathcal{H}_{eq}}{m} \sqrt{\frac{a_{eq}}{a(\tau_t)}}$.

For the tensor modes, we focus on the impact of the terms proportional to $a_0$, which are dominant for $\omega \gg \mathcal{H}$ and $\phi_0 \ll 1$. Expanding Eq.~\eqref{eq:GWequationframe} up to the order $\beta=\mathcal{H}/\omega$, we find
\begin{align}\label{eq:vofULDM}
    \frac{\mathrm{d}^2 v}{\mathrm{d}x^2} + \left( \kappa^2 - 120 \alpha^2  (\mathrm{sn}(x)^2 - 4 \mathrm{sn}(x)^6 + 4 \mathrm{sn}(x)^{10} ) + 120 \beta \alpha^2 \mathrm{cn}(x) \mathrm{sn}(x)^3 \mathrm{dn}(x) (8-11 \mathrm{sn}(x)^4) \right) v \simeq 0 ~ ,
\end{align}
where we defined for compactness
\begin{align}\label{eq:notation}
    \kappa = \frac{k}{\omega}, \qquad \alpha = \frac{\phi_0^2 \omega^2}{a^4 \Lambda^2} \equiv \frac{\alpha_\star} {(1+\beta_\star x )^{4}}, \qquad \beta = \frac{\mathcal{H}}{\omega}\equiv \frac{\beta_\star}{1+\beta_\star x} , \qquad  a(x) = a_\star \left( 1+ \beta_\star x \right) ~ .
\end{align}
In Eq.~\eqref{eq:notation} we evaluate the parameters around $\tau=\tau_\star$ which corresponds to the onset of the resonance. In this way, the subscript “$\star$” means evaluation at $\tau=\tau_\star$. We also introduced $x=\omega (\tau-\tau_\star)$ for convenience as $x\sim 0$ for $\tau\sim \tau_\star$.
Note that the leading order term in Eq.~\eqref{eq:vofULDM} scales as $\alpha_\star^2$, which corresponds to $X^2/\Lambda^4$, as we have shown in general in \S~\ref{sec:Gravitational_wave_resonances_in_DHOST}. As we will later see, the ratio between the Hubble parameter at the start of the resonance $\mathcal{H}_\star$ and the frequency $\omega$, i.e. $\beta_\star=\mathcal{H}_\star/\omega$, determines how long the mode remains in the resonance band. 

\begin{figure}
    \centering
    \includegraphics[scale=0.54]{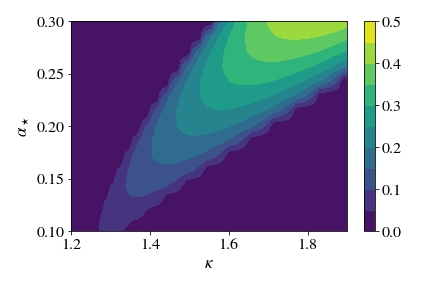} \hfill
    \includegraphics[scale=0.54]{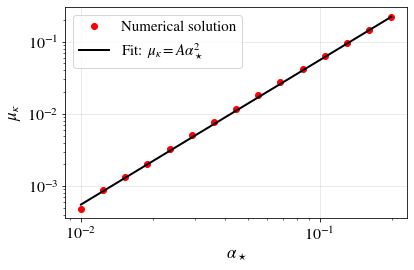}
    \caption{On the left, we plotted the Floquet exponent $\mu_\kappa$ of the first instability band without cosmic expansion ($\beta_\star=0$) in dependence on $\alpha_\star$ and $\kappa$ showing that the resonance frequency shifts towards smaller values by decreasing $\alpha_\star$. On the right, we plotted the maximal Floquet exponent $\mu_\kappa$ versus $\alpha_\star$ confirming that the resonance is in the first resonance band $\mu_\kappa = A \alpha_\star^2$. } 
    \label{fig:Floquet_exponent}
\end{figure}

We proceed with the analysis of the resonances in \eqref{eq:vofULDM}. First, since resonances occur on short time scales, we may neglect cosmic expansion ($\beta_\star=0$). In this case we have an exact Hills equations. To get the Floquet exponents we solved the Eq.~\eqref{eq:vofULDM} numerically for two independent initial conditions over one period to obtain the fundamental matrix following \cite{Amin:2014eta}. We plotted the Floquet exponents, which are given by the real part of the eigenvalues of the fundamental matrix, in Figure \ref{fig:Floquet_exponent}. On the right, we show the maximal Floquet exponent $\mu_\kappa$ versus $\alpha_\star$, which scales as $\mu_\kappa \simeq 5.55 \alpha_\star^2$ . This indicates that the resonance is in the first resonance band, similar to the standard Mathieu equation.   Then, we solved the Eq.~\eqref{eq:vofULDM} numerically and show our results in Figure \ref{fig:Floquet_exponent}. On the left, we plot the Floquet exponent $\mu_\kappa$ of the first instability band without cosmic expansion ($\beta_\star=0$) as a function of $\alpha_\star$ and $\kappa$, demonstrating that the resonance frequency shifts towards smaller values as $\alpha_\star$ decreases.

\begin{figure}[ht]
    \centering
    \includegraphics[scale=0.54]{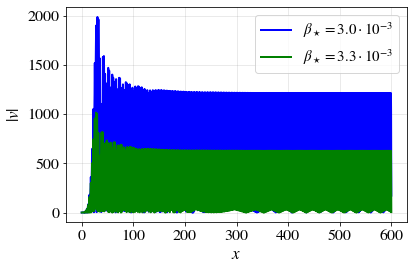}\hfill 
    \includegraphics[scale=0.54]{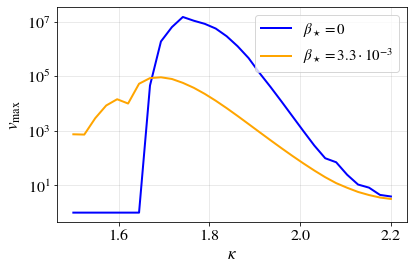} \\
    \includegraphics[scale=0.54]{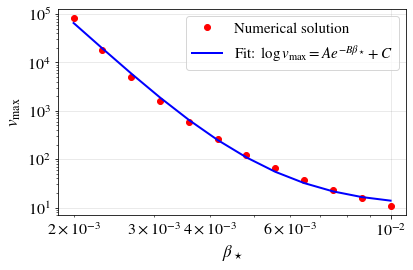} \hfill
    \includegraphics[scale=0.54]{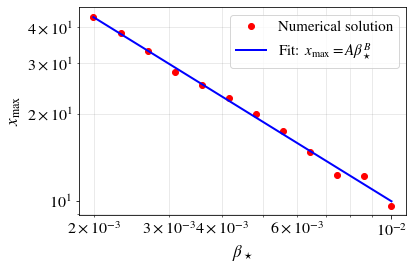}
    \caption{Impact of cosmic expansion on the parametric resonance for $\alpha_\star=0.3$. On the left upper panel, we plotted the evolution of the tensor modes for two different parameters of $\beta_\star$, showing that due to the cosmic expansion modes leave the resonance band. On the right, scale dependency for $v_{max}$ is plotted and compared without cosmic expansion up to $x_{max}$. On the left lower panel, maximal amplification $v_{\max}$ vs $\beta_\star$ is plotted showing the strong dependence on $\beta_\star$ since it fixes the timescale $x_{max}$ how long mode stays inside the resonance band, which we show on  the right lower panel.}
    \label{fig:Time_dependence_ULDM}
\end{figure}

After showing the presence of the resonant bands, let us discuss the impact of cosmic expansion. We solve Eq.~\eqref{eq:vofULDM} for the initial conditions $v(x_\star)=1,\, \mathrm{d} v/ \mathrm{d} x(x_\star)=0$ for different parameter combinations $\beta_\star$, $\kappa$ and $\alpha_\star$ and evolve it until the modes leave the resonant bands and the amplitude of $v$ remains essentially constant. The numerical results are plotted in Figure \ref{fig:Time_dependence_ULDM}. We show the numerical results in Figure \ref{fig:Time_dependence_ULDM}. On the left upper panel, we plotted the evolution of $v$ for two different values of $\beta_\star$ for $\kappa=1.81$. This figure shows that the cosmic expansion halts the resonance at a certain point, and after a slight drop, it approaches a constant value. To gain a better understanding we plotted on right upper panel the scale dependency of $v_{max}$, i.e. the maximal value of $v$ before the modes exit the resonance band at $x_{max}$. We compared this to evolving $v$ without cosmic expansion $\beta_\star=0$ for the same timescale $x_{max}$. Due to the cosmic expansion, the spectrum broadens and shifts to smaller values of $\kappa$. This shift can be easily understood since due redshifting the modes traverse through the resonance band. Therefore, we obtain the maximal amplification if the mode scans the entire band instead of starting at the center and only scanning part of the resonance band. This also explains why without cosmic expansion at the center of the resonance band, amplification is significant higher but falls off much more rapidly away from the center.

On the lower panel of Figure \ref{fig:Time_dependence_ULDM}, we study the impact of $\beta_\star$ on the maximal amplification by plotting $v_{\mathrm{max}}$ vs $\beta_\star$ on the left. The fit is an exponential decay $  \log v_{\mathrm{max}} = A e^{-B \beta_\star}+C$ with $A\simeq 22.5$, $B\simeq 479.9$ and $C\simeq 2.4$. We can see that the the resonance is highly dependent on $\beta_\star$ since it determines how long a mode stays inside the resonance band (lower right panel). The fit is given by
$x_{\mathrm{max}}= A \beta_\star^B$ with $A\simeq 0.148$ and $B\simeq-0.914$. As a very rough estimate, we have that $x_{\mathrm{max}} \beta_\star \sim \mu_\kappa$ since the width of the resonance band is of the order of the Floquet exponent. 

\subsection*{Parameter Estimate}
The maximal time span $x_\mathrm{max}$ a mode stays inside the resonance band depends on the band width, which is proportional to the Floquet exponent $\mu_\kappa$, and the relative cosmological expansion $\beta_\star$ and is at least roughly given by 
$x_\mathrm{max} \sim \mu_\kappa/\beta_\star$ as discussed in the previous section. On the other hand, the Floquet exponent in the first resonance bands scales as $\mu_\kappa \sim \alpha_\star^2$. Together, we can estimate the enhancement factor as $P \sim e^{x_\mathrm{max} \mu_\kappa}$. We can note that the enhancement is highly sensitive to $\alpha_\star$ and for a significant enhancement we need to require that $x_\mathrm{max}\mu_\kappa \sim \alpha_\star^4/\beta_\star \gtrsim 1$.
In terms of the model parameters we can express the condition as
\begin{align}
    \frac{\alpha_\star^4}{\beta_\star} \sim \left( \frac{H_\star}{\Lambda} \right)^8 \left( \frac{a_t}{a_{eq}} \right)^4 \frac{m}{H_\star} \frac{a(\tau_t)}{a_\star} \sim 10^{21} \left(\frac{m}{\mathrm{eV}}\right) \left( \frac{\mathrm{MeV}}{T_\star} \right) \left( \frac{T_{eq}}{T_t} \right)^5  \left( \frac{H_\star}{\Lambda} \right)^8 \gtrsim 1~,
\end{align}
which provides a lower bound for the mass.  
Further, we require our EFT to hold up to BBN or even earlier, which means $a_\star \leq 10^{-8}$. In a complete picture, this would entail, e.g., that the scalar field is frozen before BBN because of a very flat potential, thus rendering the modifications negligible. Only after $a_\star$ the modified gravity corrections become relevant as $X/\Lambda^2$ is sizable.

On the other hand, the resonance frequency $f \sim \omega$ is related to the mass $m$ or the coupling of the quartic interaction $\lambda$ via $\omega \sim \sqrt{\lambda} \phi_0 \sim m a_t$ which in terms of the model parameter is given by
\begin{align}
    f \sim \omega \sim 10^{11} \left( \frac{m}{\mathrm{eV}} \right) \left( \frac{a_t}{10^{-4}} \right)\,\mathrm{Hz}~,
\end{align}
or equivalently
\begin{align}
    m \sim 10^{-11} \left( \frac{10^{-4}}{a_t} \right) \left( \frac{f}{1\mathrm{Hz}} \right)\,\mathrm{eV}.
\end{align}
In order to get a feeling for the parameters let us now consider two different examples: 
\begin{itemize}
    \item LISA frequency $f_{\rm LISA} \sim 10^{-2}\,\mathrm{Hz}$
    \begin{align*}
        & m \sim 10^{-13} \left( \frac{10^{-4}}{a_t} \right)\,\mathrm{eV}~, \qquad \beta_\star \sim 10^{-7} \left( \frac{T_\star}{10\,\mathrm{MeV} } \right)^2 \left( \frac{a_\star}{10^{-8}} \right)~, \\
        & \frac{\alpha_\star^4}{\beta_\star} \sim 10^{9} \left( \frac{m}{10^{-13}\,eV} \right) \left( \frac{10\,\mathrm{MeV}}{T_\star} \right) \left( \frac{T_{eq}}{T_t} \right)^5 \left( \frac{H_\star}{\Lambda} \right)^8~.
    \end{align*}
    \item LIGO/ ET frequency $f_{\rm ET} \sim 10^2\,\mathrm{Hz}$
    \begin{align*}
        & m \sim 10^{-9} \left( \frac{10^{-4}}{a_t} \right)\,\mathrm{eV}~, \qquad \beta_\star \sim 10^{-11} \left( \frac{T_\star}{10\,\mathrm{MeV}} \right)^2 \left( \frac{a_\star}{10^{-8}} \right)~,\\
        & \frac{\alpha_\star^4}{\beta_\star} \sim 10^{13} \left( \frac{m}{10^{-9}\,\mathrm{eV}} \right) \left( \frac{10\,\mathrm{MeV}}{T_\star} \right)\left( \frac{T_{eq}}{T_t} \right)^5 \left( \frac{H_\star}{\Lambda} \right)^8
    \end{align*}
\end{itemize}
We can note that in order to have a significant enhancement in the LISA frequency range we need that the resonance starts quite close to the cutoff scale $H_\star/\Lambda \sim 1$. For the ET frequency range the parameter space is significantly larger. Note, that for $H_\star \sim 10^{-11}\,\mathrm{eV}$ depending on the frequency we may get $m \ll \Lambda$ or $m \gtrsim \Lambda$ in order to get sizeable amplifications. We do not consider the latter case as one would expect a small mass due to the breaking of the shift symmetry.

\begin{figure}
    \centering
    \includegraphics[scale=0.54]{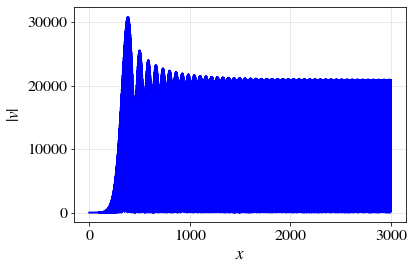}
    \includegraphics[scale=0.54]{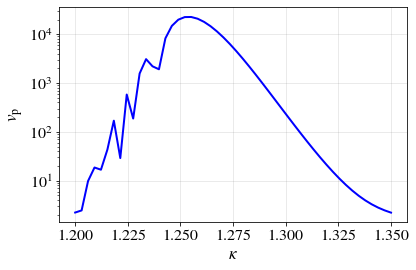} \hfill 
    \includegraphics[scale=0.54]{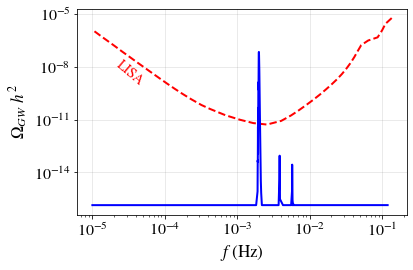} \hfill
    \includegraphics[scale=0.54]{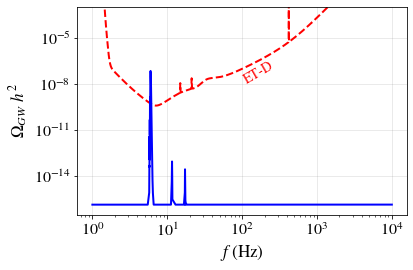}
    \caption{In the upper left panel, we plotted the evolution of the tensor modes for $\kappa=1.258 $. In the right upper panel, we depicted the scale dependency of the first resonance band after the modes reach a plateau. In the lower left panel, we illustrated the corresponding energy density $\Omega_{GW}(t_0)$ with the first three resonance bands, compared to the sensitivity curve of LISA for $\omega=10^{-2}\,\mathrm{Hz}$. In the lower right panel, we compared it to the sensitivity curve of the Einstein Telescope for $\omega=30\,\mathrm{Hz}$. The other parameters used in these plots are $\alpha_\star=0.1$ and $\beta_\star=4\cdot 10^{-4}$.}
    
    \label{fig:Time_evolution_comparison}
\end{figure}

To shorten the computation time we solve the Eq. \eqref{eq:vofULDM} for $\alpha_\star=0.1$, $\kappa =1.258$ and $\beta_\star=4\cdot 10^{-4}$ numerically which is plotted in figure \ref{fig:Time_evolution_comparison}. A value of $\alpha_\star=0.1$ is at the upper end to allow for a consistent EFT expansion and requires that $H_\star/\Lambda \sim 1$. 
We can see that the modes leave the resonance band at roughly $x_{max} \sim 400$, leading to a final amplification slightly above $2\cdot 10^4$. On the right, we plotted the spectrum of $v$ after it approaches the plateau $v_{p}$. To get an estimate for the gravitational wave spectrum nowadays $\Omega_{GW}(t_0)$, we use that the amplification is occurring on very short timescales in comparison to the Hubble parameter and after that, the tensor modes evolve like in GR so that we can estimate
\begin{align}
    \Omega_{GW}(t_0) = \Omega_{GR,GW}(t_0) v_{p}^2 ~ ,
\end{align}
where $\Omega_{GR,GW}(t_0)$ is the spectrum in standard GR for slow-roll inflation and we used the initial conditions $v(t_\star)=1$.

On the lower panel, we plotted the spectrum $\Omega_{GW}(t_0)$ in comparison to the LISA sensitivity curve with $\omega=10^{-2}\,\mathrm{Hz}$ and to the Einstein telescope with $\omega=30\,\mathrm{Hz}$ for a tensor-to-scalar ratio of $r=0.01$. Since we fix $\beta_\star$, increasing $\omega$ means that for the Einstein telescope we consider resonances which occur earlier in time. Due to the narrow resonance band, the signal is strongly peaked. We can see that for our parameter choice, the enhancement of the gravitational wave spectrum in the first resonance band is sufficient to be detected by LISA or ET, while the following two bands are below the sensitivity of these detectors. To shorten numerical computation time, we only evaluated the amplification for the first three bands. In general, there are further smaller peaks in the spectrum which are not shown in the plot since there are well below current and future detector sensitivity limits.

Lastly, let us note that the scalar field oscillations will also lead to resonances in curvature fluctuations. We can estimate the enhancement in curvature fluctuations noting that curvature fluctuations are mainly due to the dominant fluid. Since in our set-up the radiation fluid dominates the universe, curvature fluctuations are due to the radiation fluctuations, not scalar field fluctuations $\delta\phi$. In that case, scalar field oscillations will enter the equations of motion for the curvature fluctuations mainly through oscillations of the scale factor (as in the case of tensor modes). Thus, there we expect similar narrow resonances for curvature fluctuations with similar amplification factors. If so, there is the possibility that curvature fluctuations grow too much and backreact, which would lead to an upper limit on the amplification factor. However, since this depends on the details of the model and the initial amplitude of curvature fluctuations, and it is not the focus of this paper, we leave it future work. 


\section{Discussion}

We discussed how gravitational waves can be enhanced by parametric resonances due to oscillations at the background level in modified gravity theories. These resonances can originate from oscillations of the Hubble parameter, as seen in standard GR \cite{Ye:2023xyr,Alsarraj:2021yve}, but can also stem from the modified friction term and a non-trivial propagation speed of the tensor modes.
We proposed evaluating the effect in what we called the GW frame, where all effects are incorporated into the evolution of the scale factor. Our focus was on quadratic DHOST models in Class I. We demonstrated that if the modified gravity model can be purely expressed in terms of a minimal coupled scalar field plus disformal coupled radiation, the resonances are highly suppressed during the radiation domination epoch. To obtain sizeable modifications during this epoch, one needs modifications which cannot be achieved by a disformal transformation of GR. 

As a specific example, we consider ULDM with an additional non-trivial coupling between the scalar field and gravity. The evolution of the tensor modes exhibits a structure similar to a Mathieu equation, containing instability bands. The resonances are controlled by two parameters: the ratio between the resonance frequency and the Hubble parameter $\beta_\star$, which controls the time span in which the modes stay in the resonance band, and the ratio of the scalar energy density to the cutoff scale $\alpha_\star$, which controls the strength of the resonance. By tuning these parameters, the resonance can be sufficiently strong to enhance the background of primordial tensor modes from inflation to levels detectable by upcoming gravitational wave detectors such as LISA or the Einstein telescope.

Lastly, we would like to stress that the formalism developed in this work also applies to general oscillating non-minimally coupled scalar fields in the early universe, not only to the case of ULDM. Thus, finding very sharp peaks in the GW spectrum may be an indication of modifications of gravity in the early universe.

\section*{Acknowledgments}
We would like to thank Chao Chen and Zihan Zhou for helpful discussions and comments. A.G. receives support by the grant No. UMO-2021/40/C/ST9/00015 from the National Science Centre, Poland.  G.D. and A.G. are supported by the DFG under the Emmy-Noether program grant no. DO 2574/1-1, project number 496592360. The work of C.L. was supported by the grant no. 2021/42/E/ST9/00260 from the Polish National Science Centre.
J.J. was supported by the National Research Foundation of Korea (NRF) grant funded by the Korea government (MSIT) (2021R1A4A5031460). Y.F.C. is supported in part by National Key R\&D Program of China (2021YFC2203100), by CAS Young Interdisciplinary Innovation Team (JCTD-2022-20), by NSFC (12261131497, 11975020, 12005309), by 111 Project (B23042), by Fundamental Research Funds for Central Universities, by CSC Innovation Talent Funds, by USTC Fellowship for International Cooperation, by USTC Research Funds of the Double First-Class Initiative. B.W. is supported by NSFC (12003029).

\appendix

\section{Conformal coupled Matter}
\label{app:Conformal_matter}

In this appendix we explicitly show that in the case of Brans-Dicke theory, with a coupling of the type $\phi^2R$, the resonances in the GW equation stemming from the non-minimal coupling are suppressed (or similar to the minimally coupled canonical scalar field) during radiation domination. The analysis of this appendix is readily applicable to the general case of $f(\phi)R$. 
\subsection{Jordan frame}
Let us consider a simple conformal coupling of the matter sector, which in the Jordan frame is given by
\begin{align}
\label{eq:Jordan_frame_conformal}
S_J=\int \mathrm{d}^4x\sqrt{-\tilde g}\left\{\frac{1}{2}\left(1+2\xi \tilde \phi^2\right)\tilde R-\frac{1}{2}\tilde g^{\mu\nu}\partial_\mu\tilde\phi\partial_\nu\tilde \phi-\tilde V(\tilde \phi)\right\} + S_M(\tilde g_{\mu\nu},\chi) ~ .
\end{align}
The background equations of motion are given by
\begin{align}\label{FriedmannEq}
3\tilde{{\cal H}}^2=\frac{\tilde a^2 \tilde \rho_r+\tfrac{1}{2}\tilde \phi'^2+\tilde a^2 \tilde V(\tilde \phi)}{1+2\xi \tilde \phi^2+4\xi\frac{\tilde \phi'}{\tilde{{\cal H}} }\tilde \phi} ~ ,
\end{align}
and
\begin{align}
\tilde \phi ^ { \prime \prime } + 2 \tilde{\mathcal{ H }} \tilde \phi ^ { \prime } + \tilde a^2 \tilde V_{\tilde \phi}= 2 \tilde a ^ 2  \xi \tilde\phi \tilde R ~ .
\end{align}
We assume that the radiation fluid dominates, i.e., $\tilde a^2 \tilde \rho_r \gg \frac{1}{2} \tilde \phi^{\prime2} + \tilde a^2 \tilde V$, so that the Friedmann equation can be approximated as
\begin{align}
     3\tilde{{\cal H}}^2 \simeq & \frac{\tilde a^2 \tilde \rho_r}{1+2\xi \tilde \phi^2+4\xi\frac{\tilde \phi'}{\tilde{{\cal H}} }\tilde \phi} ~ .
\end{align}
In that case, in the Jordan frame, $\tilde R = \mathcal{O}(\xi)$, so that the correction to the scalar field equation of motion are of order $\mathcal{O}(\xi^2)$ and can be neglected. Furthermore, the Friedmann equation can be rewritten as
\begin{align}
     3\tilde{{\cal H}}^2 \left(  1 + \frac{4 \xi \frac{\tilde \phi^\prime \tilde \phi}{\tilde{\mathcal{H}}}}{1+ 2 \xi \tilde \phi^2}  \right)\simeq  &\frac{\tilde a^2 \tilde \rho_r}{1+2\xi \tilde \phi^2} ~ ,
\end{align}
Utilizing $\tilde \rho_r \propto \tilde a^{-4}$, we can identify a new scale factor and Hubble parameter as
\begin{align}
    a^2 = & \tilde a^2 ( 1 + 2 \xi \tilde \phi^2 )~ , \\
    {\mathcal{H}} = & \tilde{\mathcal{H}} +  \frac{2 \xi \tilde \phi^\prime \tilde \phi }{1+ 2 \xi \tilde \phi^2} ~ .
\end{align}
This is, indeed, the standard scale factor and Hubble parameter in the Einstein frame. 

\subsection{Einstein frame}
Alternatively, by performing the conformal transformation 
\begin{align}
    \tilde g_{\mu\nu} = \Omega^2 g_{\mu\nu} \qquad \Omega^{-2} =1 + 2 \xi \tilde \phi^2 ~ ,
\end{align}
we can rewrite the action \eqref{eq:Jordan_frame_conformal} in the Einstein frame 
\begin{align}
    S_E = \int \mathrm{d}^4x\, \sqrt{g} \left( \frac{1}{2} R - \frac{1}{2} g^{\mu\nu} \partial_\mu \phi \partial_\nu \phi - \Omega^4 V(\phi)   \right) + S_M(\tilde g_{\mu\nu},\chi ) ~ ,
\end{align}
where we have redefined the scalar field as
\begin{align}
    \mathrm{d}\phi = \mathrm{d}\tilde \phi \sqrt{1 + 3 \left( \frac{\Omega_{\tilde \phi}}{\Omega}\right)^2} ~ .
\end{align}
The matter sector remains unchanged due to the conformal invariance of radiation. The background equation of motions are given by
\begin{align}
    & 3 \mathcal{H}^2 = a^2 \rho_r + \frac{1}{2} \phi^{\prime 2} + a^2 V(\phi) ~ , \\
    & \phi^{\prime\prime} + 2 \mathcal{H} \phi^\prime + V_\phi =0 ~ ,
\end{align}
with $V(\phi) = \Omega^4 \tilde V  $. At leading order in $\xi$,
\begin{align}
    \mathrm{d}\phi \simeq \mathrm{d} \tilde \phi \left( 1 + 3 \xi^2 \tilde \phi^2 \right) ~ ,
\end{align}
so that as a good approximation, $\phi \simeq \tilde \phi + \mathcal{O}(\xi^2)$, and at leading order, the dynamics of the scalar field in both frames are the same. Using $\rho_r \propto a^{-4}$, the zeroth-order solution of the scale factor is $a \sim \tau$. The only leading-order contribution in $\xi$ to the Hubble parameter is inside the potential $V(\phi) \simeq (1+ 4 \xi \phi^2) \tilde V$. However, as long as the radiation fluid is dominating, this impact is additionally suppressed by $\tilde V/\rho_r$, i.e.
\begin{align}
    3 \mathcal{H}^2 \simeq a^2 \rho_r \left( 1 + \frac{\frac{1}{2} \phi^{\prime 2} + a^2 \tilde V ( 1+ 4 \xi \phi^2)}{\rho_r} \right)~ .
\end{align}

\subsection{Tensor modes}
The action for the tensor modes in both frames is given by
\begin{align}
    S_J = \int \mathrm{d}^4x\, \tilde a^2 \left( 1 + 2 \xi \tilde \phi^2 \right) \left( \tilde h_{ij}^{\prime 2} - (\partial_k \tilde h_{ij})^2 \right) ~ ,
\end{align}
and
\begin{align}
    S_E = \int \mathrm{d}^4x\, a^2  \left( h_{ij}^{\prime 2} - (\partial_k  h_{ij})^2 \right) ~ ,
\end{align}
where 
\begin{align}
    \label{eq:Scale_Factor_Jordan}
    a^2 = \tilde a^2 (1+ 2 \xi \phi^2) ~ .
\end{align}
In the Einstein frame the equation of motion for the normalized tensor modes is given by
\begin{align}
    v^{\prime\prime} + \left( k^2 - \frac{a^{\prime\prime}}{a} \right) v \approx v^{\prime\prime} + \left( k^2 + \frac{1}{6 } \frac{\phi^{\prime 2}}{a^2} + \frac{1}{3} \tilde V \left(  1 + 4 \xi \phi^2 \right)  \right) v = 0 ~ , 
\end{align}
where in the second step we have used the background equation of motions. The leading resonant effect due to the oscillating scalar field is the same as for a minimally coupled scalar field, while the modified gravity effect proportional to $\xi$ is additionally suppressed by the energy density of the scalar field as discussed in the previous section. 

In the Jordan frame the discussion is a bit more involved. Using the equation of motion
\begin{align}
    & \tilde v^{\prime\prime } + \left( k^2 - \frac{\tilde a^{\prime\prime}}{\tilde a} - 2 \xi \frac{\tilde \phi^{\prime 2} + \tilde \phi \tilde \phi^{\prime\prime} + 2 \tilde{\mathcal{H}} \tilde \phi \tilde \phi^{\prime}}{1+ 2\xi \tilde \phi^2} + 4 \xi^2 \frac{\tilde \phi^{\prime 2} \tilde \phi^2}{(1+2\xi \tilde \phi^2)^2}  \right) \tilde v =0 ~ ,
\end{align}
we might naively guess that the third term in the bracket provides a resonance which is not additionally suppressed by the energy density of the scalar field. However, using \eqref{eq:Scale_Factor_Jordan} and $\tilde \phi = \phi + \mathcal{O}(\xi^2)$ the term is exactly cancelled by $\tilde a^{\prime\prime}/\tilde a$ leading to the same conclusion as in the Einstein frame. 

Therefore, during radiation domination the non-trivial conformal coupling between the scalar field and the curvature, $\tilde \phi^2 \tilde R$, cannot enhance the parametric resonance since the contribution from the modified friction term and the background equation of motions exactly cancel each other at leading order.

\bibliography{bibliography.bib}

\end{document}